   \newcommand{\be}[0]{\begin{equation}}
   \newcommand{\ee}[0]{\end{equation}}
   \newcommand{\ba}[0]{\begin{eqnarray}}
   \newcommand{\ea}[0]{\end{eqnarray}}    
        
\documentstyle[12pt]{article}
\begin{document}
\Large
\hfill\vbox{\hbox{DTP/97/40}
            \hbox{May 1997}}
\nopagebreak

\vspace{0.75cm}
\begin{center}
\LARGE
{\bf Large-order Behaviour of the QCD Adler D-function in Planar
Approximation}
\vspace{0.6cm}
\Large

C.J. Maxwell

\vspace{0.4cm}
\large
\begin{em}
Centre for Particle Theory, University of Durham\\
South Road, Durham, DH1 3LE, England
\end{em}

\vspace{1.7cm}

\end{center}
\normalsize
\vspace{0.45cm}

\centerline{\bf Abstract}
\vspace{0.3cm}
We consider the structure of the leading ultra-violet (UV)
renormalon singularity associated with the QCD vacuum
polarization Adler D-function, in the approximation that
only planar Feynman diagrams are retained. This ``planar
approximation'' results in some simplifications, in particular
three of the four potential contributions from four-fermion
operators are shown to be absent. We are able to obtain
a fully normalized result for the leading ${n}\rightarrow{\infty}$
behaviour of the portion of perturbative coefficients
proportional to ${N_{f}^{n-r}}{N}^r$, for SU($N$) QCD with
$N_f$ quark flavours.
\newpage

The large-order behaviour of QCD perturbation series and the
structure of the renormalon singularities which give rise to it
have attracted a great deal of recent research interest 
\cite{r1}. For the QCD vacuum polarization Adler $D$-function
the leading large-order behaviour is generated by an
ultra-violet (UV) renormalon. The structure of this Borel plane
singularity can be identified following the work of Parisi
\cite{r2}, by considering the insertion of dimension-six
operators into the Green function which generates $D$ \cite{r3}.
In this paper we wish to make use of the results of ref.[3]
to study the asymptotics of the perturbative coefficients of
$D$ in the ``planar approximation'' where at each order in the
${1}/{N_f}$ expansion in SU($N$) QCD with $N_f$ quark
flavours, only the leading contribution in the large-$N$ limit
is retained. This corresponds to the evaluation of planar
Feynman diagrams \cite{r4}, and in this approximation some
simplifications occur.

The Adler D-function is defined in terms of the correlator
${\Pi}(s)$ of two vector currents in the Euclidean region,

\be
\Pi(s)(q_{\mu}q_{\nu}-g_{\mu\nu}q^{2})=4\pi^{2}i
\int\mbox{d}^{4}x\,e^{i{\bf q.x}}\langle 0|T\{J_{\mu}(x)
J_{\nu}(0)\}|0\rangle
\;,
\ee
where $s= {-q^2}>0$. By taking a derivative in $s$ one avoids an
unspecified constant and defines
\be
D(s)=-s\frac{\mbox{d}}{\mbox{d}s}\Pi(s)
\; .
\ee
In SU($N$) QCD perturbation theory,

\be
D(s)=N\sum_{f}Q_{f}^{2}\left(1+\frac{3}{4}C_{F}\tilde{D}\right)
+\left(\sum_{f}Q_{f}
\right)^{2}\tilde{\tilde{D}}
\;,
\ee
where $Q_{f}$ denotes the electric charge of the quarks and the 
summation is over the flavours accessible at a given energy.  
 The SU($N$) Casimirs are $C_{A}=N$, $C_{F}=
(N^{2}-1)/2N$.
$\tilde{\tilde D}$ denotes the contribution of ``light-by-light''
diagrams which first enter at O(${\alpha}_{s}^{3}$) in
perturbation theory.
We shall be interested in the corrections to the parton model
result, $\tilde{D}$, which in ``planar approximation'' have the 
form,

\be
\tilde{D}=\sum_{n=0}^{\infty}{d_n}{a^{n+1}}\;+
O\left(\frac{1}{N^2}\right)\;,
\ee
where ${d_0}=1$ and $a\equiv{\alpha_s}/{\pi}$ is the RG-improved
coupling.
We stress that the ${d_n}$ correspond to the perturbative 
coefficients
in ``planar approximation''. Crucially each ${d_n}$ is then a sum
of multinomials in $N_f$ and $N$ of degree $n$,

\ba
{d_n}&=&\sum_{r=0}^{n}{{N}_{f}^{n-r}}{N^r}{{d}_{n}^{[n-r]}}
\nonumber
\\
&=&\sum_{r=0}^{n}{b^{n-r}}{N^r}{{d}_{n}^{(n-r)}}
\;,
\ea
where the ${{d}_{n}^{[n-r]}}$, ${{d}_{n}^{(n-r)}}$ are pure
numbers. ${b}\equiv({11N-2{{N}_{f}}})/{6}$ is the first QCD
beta-function coefficient. The second expression is the
``$b$-expansion'' [5-7] obtained by rewriting 
${N_f}={(11N-6b)}/{2}$.

Our aim is to obtain asymptotic results for ${{d}_{n}^{[n-r]}}$
and ${{d}_{n}^{[r]}}$, for fixed $r$ and large $n$. Representing
$\tilde{D}$ as a Borel integral,

\be
\tilde{D}=\int_{0}^{\infty}\mbox{d}z\,e^{-z/a}B[\tilde{D}](z)\;.
\ee
The Borel transform $B[\tilde{D}](z)$ is expected to have a
leading UV renormalon singularity at $z=-2/b$. As noted earlier
this should receive contributions from the insertion of
dimension-six operators, so that

\be
B[\tilde{D}](z)=\sum_{i}{\frac{{K}_{i}+O(1+bz/2)}{{(1+bz/2)}
^{{\delta}_{i}}}}\;,
\ee
where the sum is over the contribution of operators ${\cal{O}}_i$
. The exponents ${\delta}_{i}$ have the form

\be
{\delta}_{i}={p_i}-{(2c/b)}+{{\lambda}_{i}}\;,
\ee
where $p_i$ is an integer, $c$ denotes the second universal
beta-function coefficient,
\be
c=\left[-\frac{7}{8}\frac{C_{A}^{2}}{b}-\frac{11}{8}\frac{C_{A}C_{F}}{b}
+\frac{5}{4}C_{A}+\frac{3}{4}C_{F}\right]\;,
\ee
and ${\lambda}_{i}$ denotes the one-loop anomalous dimension
of the operator.

In the ``planar approximation'' the residues ${K}_{i}$ and
exponents ${\delta}_{i}$ can only involve the ratio
$N/b$ in order to reproduce the multinomials ${{b}^{n-r}}{N^r}$
in eq.(5). That is we must have,

\ba
{K_i}&
=&\sum_{m}{{K}_{i}^{(m)}}\left(\frac{N}{b}\right)^{m}
\nonumber\\
{{\delta}_{i}}&=&\sum_{m}{{\delta}_{i}^{(m)}}
{\left(\frac{N}{b}\right)}^{m}\;,
\ea
where the ${K}_{i}^{(m)}$ and ${\delta}_{i}^{(m)}$ are
pure numbers.
The combination $N/b$ is the organizing parameter of an
expansion in the number of renormalon chains exchanged in
QCD planar diagrams, and replaces the $1/{N_f}$ appropriate
to the chain expansion in QED \cite{r8}.

To obtain the large-$N_f$ asymptotics ${d}_{n}^{[n]}$ we
need to consider the one-chain contribution generated by
the operator
${{\cal{O}}_1}\equiv({1/{g^4}}){{\partial}_{\nu}}{{F}^{{\nu}{\mu}}}
{{\partial}^{\rho}}{{F}_{{\rho}{\mu}}}$, with $g$ the QCD
coupling and ${{F}_{{\mu}{\nu}}}$ the field strength
tensor associated with the external vector source.
The large-$N_f$ behaviour is of course known exactly
\cite{r9,r10}. The one-loop anomalous dimension of ${\cal{O}}_1$
vanishes and 
\be
{{\delta}_{1}}=2-{(2c/b)}\;.
\ee
If we work for convenience in the ``V-scheme'' 
($\overline{\mbox{MS}}$ with scale ${{\mu}^{2}}=e^{-5/3}s$)
then one finds ${K}_{1}^{(0)}={4/9}$ \cite{r5,r9} . Exchanges
of two and more renormalon chains will involve the
four-fermion operators ${\cal{O}}_V$ and ${\cal{O}}_A$ \cite{r3,r11}
,
\ba
{{\cal{O}}_V}&\equiv&({\bar{\psi}}{\gamma_{\mu}}{T^A}{\psi})
({\bar{\psi}}{\gamma^{\mu}}{T^A}{\psi})\nonumber\\
{{\cal{O}}_A}
&\equiv&({\bar{\psi}}{\gamma_{\mu}}{\gamma_{5}}{T^A}{\psi})
({\bar{\psi}}{\gamma^{\mu}}{\gamma_{5}}{T^A}{\psi})\;,
\ea
where $T^A$ denotes colour matrices (these correspond to
${\cal{O}}_3$,${\cal{O}}_4$ in ref.[3]). The remaining four-fermion
operators which do not contain $T^A$ (${\cal{O}}_1$,${\cal{O}}_2$ in 
ref.[3])
will not contribute in the ``planar approximation'' since they
will be suppressed by powers of $N$.

The corresponding anomalous dimensions of the diagonal operator
combinations of ${\cal{O}}_V$ and ${\cal{O}}_A$ can be obtained from the
anomalous dimension matrix computed in ref.[3]. On discarding
contributions which vanish in ``planar approximation'' one
finds,

\be
{\lambda_{\pm}}=\frac{N_f}{3b}\;{\pm}
\frac{\sqrt{{16}{{N}_{f}^{2}}/{9}+{9}{N^2}}-3N}{4b}
\;.
\ee

In the large-$N$ limit the diagonal operator combinations
become ${{\cal{O}}}_{\pm}\equiv{{\cal{O}}_{V}}{\pm}{{\cal{O}}_{A}}$
. After Fierz-ing
${\cal{O}}_{-}$ is purely scalar and will not contribute to the 
vector correlator. This can be seen by showing that the
large-$N$ behaviour involves traces containing an odd number
of gamma matrices , which therefore vanish. Thus 
${K}_{-}^{(m)}=0$, $m=1,2,3,\ldots$, and there is no branch point
with exponent ${\delta}_{-}$ in ``planar approximation''.

Thus in ``planar approximation'' the relevant Borel plane 
singularities are the one-chain contribution with
branch point exponent $\delta_{1}$, and the ${\cal{O}}_{+}$ 
contribution
with exponent $\delta_{+}$. From the explicit  large-$N$
two-chain calculation of \cite{r12} one can deduce
${K}_{+}^{(1)}={-1/18}$ (in the V-scheme).     
The computation of the full  
residue ${K}_{+}$ would entail a summation of all orders
in $N/b$ , requiring exchange of any number of chains in
planar diagrams to be calculated.

The large-order behaviour resulting from the $i^{th}$ contribution
in eq.(7) is

\be
d_{n}\approx\frac{K_{i}}{\Gamma({{\delta}_{i}})}\left(-\frac{b}{2}\right)
^{n}n^{{\delta}_{i}-1}n!(1+O(1/n))\;.
\ee
The asymptotics are then determined by the operator ${\cal{O}}_i$ with   
largest ${\delta}_{i}$. In both the full theory and in
``planar approximation'' ${{\cal{O}}}_{+}$ is the dominant operator.
One has \cite{r3}
\be
{\delta}_{\pm}=3-(2c/b)+{{\lambda}_{\pm}}\;,
\ee
where ${\lambda}_{+}>0$ and ${\lambda}_{-}<0$ for ${N_f}>0$, $b>0$,
(${\lambda}_{+}$,${\lambda}_{-}$ correspond to ${\lambda}_{1}$,
${\lambda}_{4}$ tabulated in ref.[3] for various
values of $N_f$ and $N=3$). ${\cal{O}}_{-}$ decouples in ``planar
approximation'' as we have seen, and gives only a sub-leading
contribution in the full theory. Evidently from eq.(11) the
one-chain contribution is subleading by at least $O(1/n)$.
Other field operators are also sub-leading in the full theory
\cite{r3}.

Whilst the asymptotics of the full coefficient $d_n$ has an
undetermined normalization we shall now show that one can
obtain a fully normalized result for the asymptotics of the 
${d}_{n}^{[n-r]}$ coefficients in ``planar approximation''.
To derive this we begin by considering the large-$N_f$ limit
. In this limit ${\lambda}_{+}=-2$ from eq.(13) , and
${\lambda}_{1}=0$, so that the one-chain contribution
dominates by O($n$).

Expressing ${\delta}_{1}$ in eq.(11) as an expansion in
$N/b$ using eq.(9) for $c$ one finds
\be
{\delta}_{1}={2}-{\frac{13}{4}}{\frac{N}{b}}+{\frac{25}{8}}
{\frac{N^2}{b^2}}\;.
\ee
Setting ${K}_{1}^{(0)}={4/9}$ and systematically expanding   
${n}^{{\delta}_{1}-1}/{\Gamma}({{\delta}_{1}})$ in eq.(14)
in powers of $N/b$ one then obtains

\be
{d_n}\approx{\frac{4}{9}}{\biggr(-{\frac{b}{2}}\biggr)}^{n}{n}{n!}
\biggr[{1}+{\frac{N}{b}}\biggr({\frac{9}{4}}{{K}_{1}^{(1)}}
-\frac{13}{4}({\ln n}-({1-{\gamma}}))\biggr)+\ldots\biggr]
\;,
\ee
where ${\gamma}=0.57722\ldots$ is the Euler-Mascheroni constant
which arises in the expansion of ${\Gamma}({\delta}_{1})$. The
asymptotics of the sub-leading ``$b$-expansion'' coefficients
${d}_{n}^{(n-r)}$ can be obtained from successive terms in
the square bracket in eq.(17). Their evaluation would 
require the calculation of the ${K}_{1}^{(m)}$ for ${m}\leq {r}$,
that is the insertion of the operator ${\cal{O}}_{1}$ into
planar diagrams with at least $r+1$ renormalon chains.

The large-order behaviour of the ${d}_{n}^{[n-r]}$ coefficients
on the other hand is determined by the ``leading-$b$'' $b^n$ term
in eq.(17). Expanding this using the binomial theorem gives a
contribution ${n^r}{N^r}{{N}_{f}^{n-r}}/{r!}$, which will
dominate by at least O($n$) over the contributions arising
from the ${({N/b})}^{m}$ terms in the square bracket. Thus
one obtains the fully normalized asymptotic behaviour
(in the V-scheme)

\be
{{d}_{n}^{[n-r]}}\approx{\frac{4}{9}}{{n}^{r+1}}{\frac{n!}{r!}}
{{6}^{-n}}{\left(-{\frac{11}{2}}\right)}^{r}
{\left(1+O\left({\frac{1}{n}}\right)
\right)}\;.
\ee

Notice that for eq.(18) to hold it is crucial that, as argued
above, there is no ${\cal{O}}_{-}$ contribution in ``planar
approximation''. In the large-$N_f$ limit ${\lambda}_{-}=0$
and since ${\cal{O}}_{-}$ first potentially contributes at
the two-chain level , and is thus O(${1}/{N_f}$) suppressed
relative to the one-chain contribution, both should contribute
equally to the asymptotics of the sub-leading ${d}_{n}^{[n-r]}$
, $r>0$. Thus if ${K}_{-}^{(1)}$ were not to vanish, the $4/9$
in eq.(18) would be replaced by $\left({\frac{4}{9}}+
{\frac{{{K}_{-}^{(1)}}{r}}{2}}\right)$. The 
result in eq.(18) was derived in ref.[5] assuming only the
one-chain contribution to the leading UV renormalon 
singularity. We have shown here that it remains valid on
including any number of chains.

We now consider the large-$N$ limit and the asymptotics of
${d}_{n}^{[r]}$. In this limit ${\cal{O}}_{+}$ is the
dominant operator and ${\lambda}_{+}=0$ from eq.(13).
Thus the one-chain contribution is sub-leading by
O($1/n$).
Returning to eq.(5) we note that it can be rewritten in yet
a third form,

\be
{d_n}=\sum_{r=0}^{\infty}{{b}^{n-r}}{{N}_{f}^{r}}{{d}_{n}^
{<n-r>}}\;,
\ee
so we can obtain an alternative ``$b$-expansion'' by
writing $N=(6b+2{N_f})/11$. Correspondingly we can write the
residues $K_i$ and exponents ${\delta}_{i}$ in eq.(7) as
expansions in ${N_f}/b$.
\ba
{K_i}&=&\sum_{m}{\hat{K}_{i}^{(m)}}{\left({\frac{N_f}{b}}\right)}
^{m}\nonumber\\
{{\delta}_{i}}&=&\sum_{m}{\hat{\delta}_{i}^{(m)}}
{\left({\frac{N_f}{b}}\right)}^{m}\;,
\ea
where the hat distinguishes these coefficients from the $N/b$
expansion in eq.(10). Whilst the $N/b$ expansion corresponds
directly to evaluating planar diagrams with an increasing
number of renormalon chains, the ${N_f}/b$ expansion has no
such direct diagrammatic interpretation. Evaluation of
$\hat{K}_{i}^{(m)}$ requires an all-orders resummation of
the $N/b$ expansion. For instance
\be
{\hat{K}_{+}^{(0)}}=\sum_{m=1}^{\infty}{{K}_{+}^{(m)}}
{\left({\frac{6}{11}}\right)}^{m}\;.
\ee
As noted earlier from the exact large-$N$ two-chain
calculation of ref.[12] one can infer ${K}_{+}^{(1)}={-1/18}$,
so with our present knowledge $\hat{K}_{+}^{(0)}={(-1/33)}+\ldots$
, it is of course not even guaranteed that the sum in eq.(21)
converges.

On expanding eq.(13),and eq.(9) for $c$, in ${N_f}/b$ one
finds

\be
{\delta}_{+}=\frac{261}{121}+\frac{263}{726}{\frac{N_f}{b}}+
\frac{3680}{9801}{\frac{{N}_{f}^{2}}{b^2}}+\ldots\;.
\ee
Inserting these ${N_f}/b$ expansions into eq.(14) gives,
analogous to eq.(17),

\ba
{d_n}&\approx&\frac{{\hat{K}_{+}^{(0)}}{(-b/2)}
^{n}{{n}^{140/121}}{n!}}{{\Gamma}({261/121})}
\biggr[1+\frac{N_f}{b}\biggr(\frac{\hat{K}_{+}^{(1)}}
{\hat{K}_{+}^{(0)}}+\frac{263}{726}({\ln n}
-{\psi}({{261}\over{121}}))\biggr)+\ldots\biggr]\nonumber\\
& &
\ea
where ${\psi}(x)$ is the logarithmic derivative of ${\Gamma}(x)$
which arises in the expansion of ${\Gamma}({\delta}_{+})$
. In principle the asymptotic behaviour of the ${d}_{n}^{<n-r>}$
coefficients in the alternative ``$b$-expansion'' in eq.(19)
can be derived from the successive terms in the square
bracket in eq.(23). However, as pointed out above, the
evaluation of the $\hat{K}_{+}^{(m)}$ requires a resummation
of any number of renormalon chains.

The large-order behaviour of the ${d}_{n}^{[r]}$ coefficients
will again be determined by the ``leading-$b$'' $b^n$ term
in eq.(23). On expansion using the binomial theorem one
obtains a contribution ${n}^{r}{N}_{f}^{r}{N}^{n-r}/r!$ , which
will dominate by at least O($n$) over the ${({N_f}/b)}^{m}$
terms in the square bracket. So that the leading large-order
behaviour is

\be
{{d}_{n}^{[r]}}\approx{\frac{\hat{K}_{+}^{(0)}}{{\Gamma}\left(
{261/121}\right)}}{{n}^{r+140/121}}{\frac{n!}{r!}}
{\left(-{\frac{11}{12}}\right)}^{n}{\left(-{\frac{2}{11}}\right)}
^{r}{\left(1+O\left({\frac{1}{n}}\right)\right)}\;.
\ee

Whilst the evaluation of $\hat{K}_{+}^{(0)}$
would require the all-chains resummation of eq.(21) by taking
ratios  we can arrive at the fully normalized result

\be
\frac{{d}_{n}^{[r]}}{{d}_{n}^{[s]}}\approx{{n}^{r-s}}
{\frac{s!}{r!}}{\left(-{\frac{2}{11}}\right)}^{r-s}
{\left(1+O\left({\frac{1}{n}}\right)\right)}\;.
\ee
Similarly from eq.(18) for ${d}_{n}^{[n-r]}$ one can obtain

\be
\frac{{d}_{n}^{[n-r]}}{{d}_{n}^{[n-s]}}\approx{{n}^{r-s}}
{\frac{s!}{r!}}{\left(-{\frac{11}{2}}\right)}^{r-s}
{\left(1+O\left({\frac{1}{n}}\right)\right)}\;.
\ee

The results in eqs.(25),(26) hold universally for any QCD
observable whose large-$N$ and large-${N_f}$ asymptotics
are each dominated by a {\it single} branch-point (operator)
in the Borel plane. For instance in the present case of the
vector correlator we have seen that the large-$N$ asymptotics
are dominated by ${\cal{O}}_{+}$ ,and the large-${N_f}$
asymptotics by the one-chain contribution. This ensures the  
weak dominance of the large-$N$ and large-${N_f}$ derived
``leading-$b$'' terms evident in eqs.(18) and (24), that is on  
binomial expansion they generate the correct leading
asymptotics for ${d}_{n}^{[r]}$ and ${d}_{n}^{[n-r]}$.

An amusing speculation is that if one were able to perform
the all-chains summation in eq.(21) one would find
$\hat{K}_{+}^{(0)}=0$, then assuming that the
O($1+bz/2$) term in the numerator of the ${\delta}_{+}$
contribution also vanished in the large-$N$ limit,
the leading large-$N$ asymptotics would in fact come from
the one-chain contribution. If the two and higher chain
contributions to the residue $K_{1}$ also summed to zero
in the large-$N$ limit,

\be
\sum_{m=1}^{\infty}{{K}_{1}^{(m)}}{\left({\frac{6}{11}}\right)}
^{m}=0\:,
\ee
then the standard large-${N_f}$ derived ``leading-$b$'' term
,which is clearly exact in the large-${N_f}$ limit , would
also give the correct leading asymptotic behaviour in the
large-$N$ limit , up to ${n}^{-2c/b}$ effects dependent on
the second beta-function coefficient which vanish in
large-${N_f}$ but contribute in large-$N$. This would give
the ``leading-$b$'' approximation [5-7], which has found
extensive phenomenological application [5,7,13-15], a
much stronger motivation.

We stress that this is pure speculation. The only fact at our
disposal is the  two-chain approximation to $\hat{K}_{+}^{(0)}$
derived from the calculation of ref.[12],
$\hat{K}_{+}^{(0)}\approx{-1/33}$. This small and negative
value makes it not inconceivable that ${K}_{+}$ has a zero
at ${N_f}=0$ (equivalent to the large-$N$ limit
${N/b}={6/11}$ in ``planar approximation'') and is thereafter
positive for ${N_f}>0$ in asymptotically-free QCD.

\section*{Acknowledgements}

We would like to thank Martin Beneke for a number of
stimulating and informative discussions which have crucially 
helped in the development of this work. The CERN TH Division
is thanked for its hospitality during the period when this
work was completed.


\newpage
\begin{thebibliography}{99}
\bibitem{r1} For a recent review see: Jan Fischer,
``On the Role of Power Expansions in Quantum Field Theory'',
PRA-HEP 97/06 [hep-ph/9704351].
\bibitem{r2} G.Parisi, Phys. Lett. {\bf B76} (1978) 65.
\bibitem{r3} M.Beneke,V.M.Braun and N.Kivel,``Large-order
Behaviour due to Ultraviolet Renormalons in QCD'',
CERN-TH/97-50 [hep-ph/9703389].
\bibitem{r4} G.`t Hooft ,Commun. Math. Phys. 86 (1982) 449.
\bibitem{r5} C.N.Lovett-Turner and C.J.Maxwell , Nucl. Phys.
{\bf B432} (1994) 147 ; ibid {\bf B452} (1995) 188.
\bibitem{r6} D.Broadhurst and A.G.Grozin, Phys. Rev. {\bf D52}
(1995) 4082.
\bibitem{r7} M.Beneke and V.M.Braun, Phys. Lett. {\bf B348}
(1995) 513.
\bibitem{r8} M.Beneke and V.A.Smirnov, Nucl. Phys. {\bf B472}
(1996) 529.
\bibitem{r9} M.Beneke, Nucl. Phys. {\bf B405} (1993) 424.
\bibitem{r10} D.J.Broadhurst, Z. Phys. {\bf C58} (1993) 339.
\bibitem{r11} A.I.Vainshtein and V.I. Zakharov, Phys. Rev. Lett.
73 (1994) 1207 [Erratum: ibid 75 (1995) 3588]; Phys. Rev.       
{\bf D54} (1996) 4039.
\bibitem{r12} S.Peris and E.de Rafael, ``Low Energy QCD and
Ultraviolet Renormalons'' [hep-ph/9701418].
\bibitem{r13} P.Ball,M.Beneke and V.M.Braun, Nucl. Phys.
{\bf B452} (1995) 563.
\bibitem{r14} M.Neubert, Nucl. Phys. {\bf B463} (1996) 511.
\bibitem{r15} C.J.Maxwell and D.G.Tonge, Nucl. Phys. {\bf B481}
(1996) 681.
\end{thebibliography}
\end{document}